\documentclass[times, review, 11pt]{elsarticle}
\usepackage[utf8]{inputenc}

\usepackage[english]{babel}
\usepackage[top =2cm, bottom=2cm, left=2cm, right=2cm]{geometry}

\usepackage{amsmath, amssymb, amsthm, amsfonts}
\usepackage{graphicx, epstopdf, xcolor, layout}
\usepackage{booktabs}

\usepackage{microtype}
\usepackage{url}

\biboptions{sort&compress}
\graphicspath{{images/}}

\usepackage{soul}

\soulregister\ref7
\soulregister\pageref7
\soulregister\cite7
\soulregister\citet7
\soulregister\citep7

\begin{document}

\begin{frontmatter}

\journal{Omega}
\title{Seasonal hydrogen storage decisions under constrained electricity distribution capacity}

\author[]{Jan Eise Fokkema\corref{corrauthor}} \ead{j.e.fokkema@rug.nl} 
\author[]{Michiel A.\ J.\ uit het Broek} \ead{a.j.uit.het.broek@rug.nl} 
\author[]{Albert H.\ Schrotenboer} \ead{a.h.schrotenboer@rug.nl} 
\author[]{Martin J.\ Land} \ead{m.j.land@rug.nl} 
\author[]{Nicky D.\ Van Foreest} \ead{n.d.foreest@rug.nl} 

\cortext[corrauthor]{Corresponding author: Nettelbosje 2, 9747 AE Groningen, the Netherlands.}

\address{Department of Operations, University of Groningen, PO Box 800, 9700 AV Groningen, the Netherlands}

\begin{abstract}
\small
The transition to renewable energy systems causes increased decentralization of the energy supply. Solar parks are built to increase renewable energy penetration and to supply local communities that become increasingly self-sufficient. These parks are generally installed in rural areas where electricity grid distribution capacity is limited. This causes the produced energy to create grid congestion. Temporary storage can be a solution. In addition to batteries, which are most suitable for intraday storage, hydrogen provides a long-term storage option and can be used to overcome seasonal mismatches in supply and demand. In this paper, we examine the operational decisions related to storing energy using hydrogen, and buying from or selling to the grid considering grid capacity limitations. We model the problem as a Markov decision process taking into account seasonal production and demand patterns, uncertain solar energy generation, and local electricity prices. We show that ignoring seasonal demand and production patterns is suboptimal. In addition, we show that the introduction of a hydrogen storage facility for a solar farm in rural areas may lead to positive profits, whereas this is loss-making without storage facilities. 
In a sensitivity analysis, we show that only if distribution capacity is too small, hydrogen storage does not lead to profits and reduced congestion at the cable connection. When the distribution capacity is constrained, a higher storage capacity leads to more buying-related actions from the electricity grid to prevent future shortages and to exploit price differences. This leads to more congestion at the connected cable and is an important insight for policy-makers and net-operators.

\end{abstract}


\end{frontmatter}

\section{Introduction}

Renewable energy sources have become increasingly popular. For example, renewable energy production in the EU has increased from 9.6\% in 2004 to 18.9\% in 2018 \citep{RES}. However, seasonality mismatches between supply and demand are one of the main challenges that should be dealt with to facilitate growth in renewable energy production. Large solar parks tend to be located in rural areas where land is relatively cheap, even though the electricity grid infrastructure is often limited. This typically causes cable congestion at the location where the solar park is connected to the grid, which in turn causes outages, grid balance problems, and affects operational costs of the electricity grid \citep{vargas2014wind, KUMAR2005153}. The high peaks of solar energy production in summer often cause cable congestion which may inhibit the installation of solar parks.  Designing electricity grids that can accommodate location-specific energy generation and take into account physical constraints is therefore a challenging task \cite{MARKLEHU2020102071}. 

Hydrogen storage is a useful flexibility option and can spread the feed-in to the electricity grid throughout the year, thereby mitigating cable congestion and bridging seasonality gaps in demand and production \citep{alanne2017zero}. Storage is one of many flexibility-related options that also include market mechanisms for power re-dispatch, curtailment, conversion, and demand-side management \citep{KONDZIELLA201610, Kamga2009, vargas2014wind}.

Storage owners connected to an external electricity grid face variable electricity prices as a result of market mechanisms to balance supply and demand, and to reduce congestion. For example, location-specific (nodal) electricity prices can be an effective solution to reduce grid congestion caused by renewable energy sources \citep{PAPAEFTHYMIOU201669}.  Furthermore, time-of-use tariffs are expected to become more common since these provide advantages to grid operators in alleviating expected congestion \cite{SOARES2020102027}.  The resulting market mechanisms facilitate the re-dispatching of energy production by stimulating additional production in areas without congestion and reducing production in areas of congestion by using nodal prices \citep{wang2017impact}. However, this requires storage owners to take into account these market mechanisms to maximize their profits. 

Electricity generated by solar parks can be fed to an external electricity grid, used to directly supply a local electricity demand of households at a nearby location or stored in a nearby storage location. The local consumption of the generated electricity avoids the transportation of energy over long distances and enables fulfilling a local electricity demand.  Electricity demand of consumers is also characterized by differences in seasonality \cite{ARORA201647}.  The confinement of the produced energy in the area of generation, seasonality differences in supply and local demand, and the presence of variable nodal electricity prices have important consequences for the decision to store energy. Efficient operational strategies that determine when energy is stored, sold, or bought from the electricity grid are key in achieving successful renewable energy penetration. Hydrogen storage is characterized by relatively high investment costs and limited conversion efficiencies. Therefore, strategies related to storage and interaction with the electricity grid become even more important in enhancing the economic viability of hydrogen storage as a flexibility option, that is especially suitable for long-term seasonal storage.  

In this study, we examine the operational problem of a hydrogen storage owner who needs to decide how much to store, buy from, and sell to the electricity grid throughout the year. This decision is affected by the presence of seasonality in supply and demand. This decision is dependent on the level of solar energy production, the amount of local electricity demand, the amount of hydrogen in storage, and the current electricity price. Additionally, solar energy production levels, electricity demand, and prices in the future are uncertain. For example, even though solar energy production can be predicted rather accurately for several days in advance, solar energy production levels for specific days are uncertain when predicted for longer periods such as months. As a result of seasonal patterns, the stochastic behavior of solar energy production and electricity demand is time-dependent. These aspects affect storage decisions throughout the year. A Markov decision process formulation is proposed to obtain optimal policies for the above problem. 

Our main contributions are as follows. In an extensive numerical study, we show how hydrogen storage can either solve or cause congestion problems at the cable to which the solar park is connected as a result of selling or buying-related decisions. Secondly, we identify the characteristics of optimal storage policies that underlie the above-mentioned issues. These characteristics include price thresholds for each period which depend on the inventory level, price, and net production after demand. We show policies in summer and winter by taking into account seasonality related to supply and demand. This includes the actions which are taken during overages and shortages throughout the year. Thirdly, we indicate how different combinations of storage and distribution capacity affect these decisions. Next, we analyze profits, congestion levels, and electrolyzer utilization. Finally, we show how conversion losses and differences between selling and buying prices affect these results. 

Results show that optimal policies are characterized by price thresholds which separate different types of actions. These include buying the maximum possible quantity, selling exact overages or buying exact shortages, storing overages or obtaining shortages from storage, or selling the maximum amount possible. When distribution capacity is constrained, congestion at the cable connection is mostly caused by buying-related actions in winter, which are used to cover potential future shortages. For higher levels of distribution capacity, congestion is mostly caused by selling-related actions of the overages in the summer. While it may be expected that storage enables reducing congestion, increased levels of storage capacity have an adverse affect on congestion. This is because storage enables increasing buying-related actions to prevent future shortages and exploiting price differences. Results also indicate that a lower electrolyzer utilization as a result of a large capacity is associated with higher profits than a low electrolyzer capacity with a higher utilization rate. This indicates that a high utilization level of the electrolyzer is not necessarily an indication of improved feasibility.


The remainder of this paper is organized as follows. A literature review is presented in Section~\ref{sec:literature}. Section~\ref{sec:problem} describes the problem and Section \ref{sec:DP} formulates a model. Section~\ref{sec:numeric} provides an overview of the calibration of the parameter settings and the base case system that we consider. Section~\ref{sec:sensitivity} provides a sensitivity analysis of key performance indicators based on the parameter settings of the capacities of each of the system elements.  Section~\ref{section: conclusion} provides concluding remarks. 

\section{Literature review}
\label{sec:literature}

 The existing literature has mostly addressed energy management strategies in which the owner of an energy storage device decides when to buy or sell energy from or to the grid. For a detailed review of energy management decisions for electric storage systems, we refer to \cite{WEITZEL2018582} and \citep{ZAKARIA20201543}. While seasonality differences between supply and demand are important characteristics of renewable energy systems, most papers focus on intraday and day-ahead buying and selling decisions using battery storage for short planning horizons and small discretization levels. In contrast, our approach focuses on hydrogen storage decisions for longer planning horizons covering seasonality during a year. We take into account limited electricity grid distribution capacity which support the need for local storage. Seasonal patterns in supply and demand, limited grid infrastructure, and electrolyzer and fuel-cell constraints all have consequences for storage decisions.


The literature on energy management decisions optimizes grid interactions using storage by treating supply levels as deterministic \citep{dufo2015optimisation, zhang2017comparative, Pelzer2016} or by optimizing the buying, selling or storage decisions in which the uncertainty of the generated renewable energy is taken into account \citep{Jiang2015, jiang2015approximate, shin2017operational, grillo2015optimal, hannah2011approximate, HASSLER2017199, gonsch2016sell, Keerthisinghe2019}.  Literature on energy procurement decisions without storage include \citet{WANG2019212} and \citet{WOO200670}. For example, \citet{WANG2019212} analyze an energy procurement problem for a centralized energy aggregator who can control both procurement and consumption within a 24 hour planning horizon in which wind energy is generated. From the perspective of a grid distribution operator, \citet{WOO200670} address the energy procurement decisions of a grid distributor that need to balance procurement risks and expected costs. Regarding prices, \citet{DENSING2013321, Zhou2019, Jiang2015, jiang2015approximate, HASSLER2017199} specifically take into account stochastic electricity prices, whereas \citet{Keerthisinghe2019, STEFFEN2016308, grillo2015optimal, shin2017operational} treat these as deterministic.  These studies are explained in more detail below. 

Studies that optimize the use of storage with buying or selling from or to the grid mostly focus on detailed decisions within one day \citep{Jiang2015, jiang2015approximate, shin2017operational, grillo2015optimal, hannah2011approximate, HASSLER2017199, gonsch2016sell, Keerthisinghe2019}. For example, \citet{Jiang2015} address the arbitrage problem with energy storage to place bids in an hour-ahead spot market. \citet{grillo2015optimal} optimally schedule batteries with renewable energy. \citet{HASSLER2017199, gonsch2016sell} optimize energy arbitrage decisions for short time horizons within one day and time intervals of 15 minutes. \citet{Keerthisinghe2019} develop energy-storing policies for a battery in a residential home within single days. In contrast, \citet{shin2017operational} have addressed the problem for both intraday and yearly planning horizons. \citet{Zhou2019} address the energy storage arbitrage problem within a week for 5-minute periods. They address seasonality by using specific parameter settings for each week that is solved. While these papers all address detailed arbitrage and storing decisions for batteries and short time horizons using batteries, none of them consider hydrogen which is seen as a more viable option for long-term storage. Most papers also do not provide an integrated approach to address the issue of seasonality over one year.

Related literature on energy storage and arbitrage which has included transmission or distribution capacity constraints is relatively scarce. \citet{FERTIG20112330} investigate the economics of pairing a wind farm with compressed air energy storage and limited transmission capacity. They use heuristic control policies to decide on buying and selling to the grid. Most work that incorporates transmission constraints focuses on energy storage planning from a strategic perspective rather than an operational perspective. For example, \citet{BABROWSKI2016228} optimize storage planning for the German electricity sector while including transmission constraints. \citet{wang2017impact} examine to what extent transmission congestion affects the profitability of arbitrage by energy storage including transmission constraints. \citet{jorgenson2018analyzing} analyze to what extent transmission or storage can assist in reducing curtailment.

To the best of our knowledge, \citet{Zhou2019} and \citet{gonsch2016sell} are the only papers that have included transmission constraints in focusing on the operational decision of when to buy, sell or store. They have addressed wind-based electricity with co-located storage. However, their numerical study encompasses only one week and does not consider hydrogen storage which would become relevant for longer terms. To the best of our knowledge, the seasonality effects related to production, demand, and nodal prices of large-scale renewable energy generation and storage have not yet been considered.



 



\section{Problem description}
\label{sec:problem}

\begin{figure}[ht!]
	\centering

	\resizebox{300pt}{!}{%
		\includegraphics[]{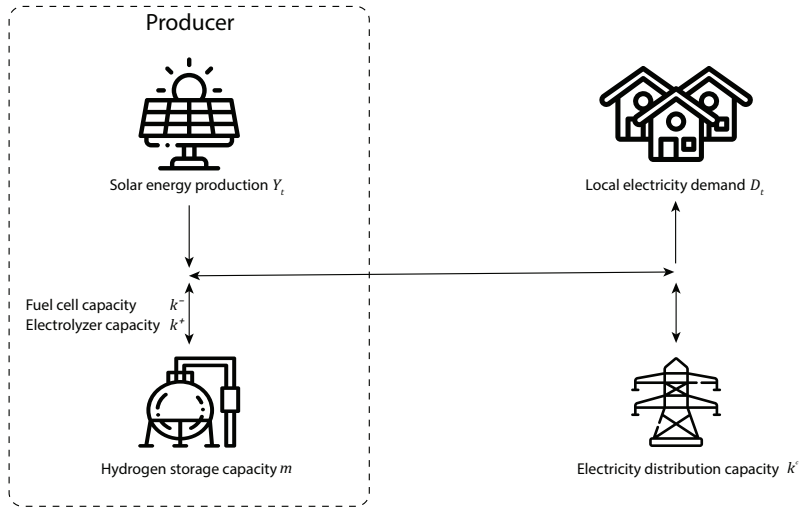}}
	
	\caption{Visual representation of the studied system}
	\label{figure:system}
\end{figure}

We consider a profit-maximizing renewable energy producer using a photovoltaic (PV) system (i.e., solar panels) who is also the storage owner and is responsible for satisfying local electricity demand. For instance, it may form a self-sufficient community together with a small village. The energy producer can also sell or buy electricity from an electricity grid, and a co-located hydrogen storage is used to temporarily store electricity in the form of hydrogen. We consider a time horizon $\mathcal T$ that resembles a complete year, and each period $t \in \mathcal T$ resembles a single day. Figure~\ref{figure:system} provides a graphical overview of the considered system \cite{flaticon}. The left side of Figure~\ref{figure:system} shows the solar energy producer and the hydrogen storage facility and the right side shows the local electricity demand and electricity grid connection. Our goal is to decide upon when and how to 1) sell and buy electricity from the grid, 2) store or consume hydrogen from our local storage to satisfy local demand and maximize profits.  We assume the owner of the storage and PV facilities, and the households are connected to the grid through an external connection and are the single users of this connection. 

In the following, we describe our system in detail. Table~\ref{table:params} provides an overview of all the parameters and variables.

The installed capacity of solar energy production in MWp is assumed constant throughout the year and denoted by $w$. Solar energy production per day is a random variable $Y_t$, where the dependency on the period follows from seasonal differences in energy production throughout the year. Local demand for electricity is denoted by the random variable $D_t$ and is assumed to be normally distributed with mean $\mu_t$ and a period-independent standard deviation $\sigma$. 

As it is optimal to always satisfy local demand with local supply of solar energy (this comes at no cost), we assume that the produced electricity $Y_t$ is first used to supply the local demand $D_t$. We then convert our production and demand process in a net solar energy production level $\bar{Y}_t := Y_t - D_t$. 
The net solar energy production per day $\bar{Y}_t$ can then be modeled via a truncated probability distribution $f^{\bar{y}}(t)$ with a maximum of $l^+_t$ in period $t$ due to a restricted installed capacity of solar energy. 

Hydrogen inventory~$x_t$ is held inside a hydrogen storage tank with energy capacity~$m$ and is filled using an electrolyzer with a maximum rate~$k^+$ at which energy can be stored per period and a conversion efficiency of~$\alpha$ where $\alpha \leq 1$. Moreover, the producer can obtain at most~$k^-$ energy units per period from storage due to a limited fuel cell capacity. The maximum amount of electricity that can be sold to or bought from the grid at the end of a period is denoted by the grid distribution capacity $k^c$. Throughout this paper, capacities (with the exception of installed solar energy capacity) are defined as the maximum amount of energy per period that can be handled by the related component in our system. 

Electricity prices are stochastic and are modeled as an AR(1) process via $C_t = \theta C_{t-1} + \xi_t$, where $\xi_t \sim N(0, \sigma^c)$. Similar to \citep{DENSING2013321, Zhou2019}, we assume that energy producer and storage owner is sufficiently small so that it is a price taker that cannot influence prices. Consequently, solar energy can be sold to the grid at a price of $C_t$. The producer can also buy from the grid at price $C_t + c^+$, where $c^+ \geq 0$ is a fixed price markup. As is commonly assumed, it is not possible to simultaneously buy and sell from the grid \citep[see, e.g.,][]{Zhou2019}.

The solar energy producer makes the buying, selling and storing decisions at the end of each period $t \in \mathcal T$. Since hydrogen conversion is associated with relatively high conversion losses and electrolyzers perform better at stable loads, we assume that intraday load fluctuations are handled by a battery. Since our focus is on seasonal differences on an annual time horizon, conversion losses of batteries are limited to 82\% and 90\% \cite{Telaretti2016} and the modeling of intraday battery load fluctuations are considered outside the scope of this paper.      

The objective of the storage owner is to maximize the expected future profits related to interacting with the electricity grid during the planning horizon. The decisions made in each period affect the total profit. These decisions have to satisfy several detailed constraints and depend on the state of the system at the end
of a period. We discuss these aspects in the next section. 

\begin{table*}[ht!]
		\small
		\centering
		
		\caption{Sets, parameters and state variables}
		\label{table:params}
		
		\begin{tabular}{ll}
		    \toprule
		    Sets&\\
		    \midrule
		    $\mathcal T$     & Set on the number of periods $\mathcal T = \{0, \ldots, T\}$\\
			\midrule
		    Parameters&\\
			\midrule
			$w$ &  Installed peak capacity of the solar park (MWp)\\
			$l^+_t$ & Maximum amount of solar energy that can be generated (MWh) in period $t$\\
			$m$ & Maximum hydrogen inventory level (storage capacity, MWh)\\
			$k^c$ &  Maximum load sent to the grid per period (distribution capacity, MW)\\
			$k^+$ & Maximum load at which energy can be stored per period (electrolyzer capacity, MWh)\\    
			$k^-$   & Maximum load from storage to electricity per period (fuel-cell capacity, MWh)\\
			$c^+$ & Price markup added to the selling price for buying energy from the grid \\
			$\alpha$ & Conversion efficiency to storage\\
			$s$ & Penalty per unit of unmet demand\\
			\midrule
			State variables&\\
			\midrule
			$\bar{y}_t$ & Net production realization after demand (MWh) in period $t$\\
			$c_t$  & Prevailing selling price of electricity in period $t$\\
			$x_t$     & Inventory level (MWh) in period $t$\\
			\midrule
			Stochastic variables&\\
			\midrule
			$Y_t$ & Solar energy production in period $t$\\
			$\bar{Y}_t$ & Net production level after demand in period $t$\\
			$C_t$ & Electricity prices in the local spot market in period $t$\\
			$D_t$ & Local electricity demand in period $t$ \\
			\bottomrule
		\end{tabular}
\end{table*}

\section{Markov Decision Process Formulation}
\label{sec:DP} 

We formulate our problem as a Markov decision process (MDP). We first describe our state and action spaces, and the constraints upon them. We also specify our reward function. We then discuss how we discretized our state and action spaces, and formally define our MDP which we solve via backward dynamic programming.

\subsection{State and action space}

At the end of period $t$, we observe an inventory level $x_t$, price level $c_t$ and net production level after demand $\bar{y}_t$. Let $S_t(\bar{y}_n, x_t, c_t, c_{t-1}) \in \mathcal{S}$ be the state of our system in period $t$. We write $\mathcal{S}$ for the state space. For each state $S_t \in \mathcal{S}$, we define the action $u(S_t) \in \mathbb{R}$ as the amount of energy units to buy from or sell to the grid at the end of period $t$. Negative values represent the amount of energy units to buy from the grid.

The action $u$ is bounded by the characteristics of the state $S_t$. It is most easily described if we consider the range of actions $[-u^{\min}(S_t), u^{\max}(S_t)]$, where $u^{\min}(S_t) \geq 0$ denotes the maximum amount of energy bought from the grid at the end of a period, and $u^{\max}(S_t) \geq 0$ denotes the maximum amount of energy that can be sold to the grid at the end of a period. In the following, we describe for each state $S_t \in \mathcal{S}$ how to obtain $u^{\min}(S_t)$ and $u^{\max}(S_t)$.

The maximum amount of energy that can be bought $u^{\min}(S_t)$, for all $S_t \in \mathcal{S}$ is the largest value which must satisfy three constraints such that
\begin{align}
    u^{\min}(S_t) \leq k^c, \quad u^{\min}(S_t) \leq (m - x_t - \alpha \bar{y}_t)/\alpha, \quad u^{\min}(S_t) \leq k^+ - \bar{y}_t.
\end{align}

The first indicates that the action should not exceed the maximum amount bought from the grid per period. The second indicates that the maximum inventory level should be considered. The third indicates that the maximum amount obtained from storage per period should be respected. 
The maximum amount of energy that can be sold $u^{\max}(S_t)$, for all $S_t \in \mathcal{S}$ is the largest value such that

\begin{align}
   u^{\max}(S_t) \leq k^c, \quad u^{\max}(S_t) \leq x_t + \bar{y}_t  \quad u^{\max}(S_t) \leq k^-,
\end{align}
which indicates that the cable distribution capacity should be respected, that we can sell at most the inventory we have plus the net overage, and that the fuel cell capacity should be respected. Note that these actions allow for unmet demand, in case $\bar{y}_t < 0$. We, therefore, introduce a penalty $s$ per unit of unmet demand that is set sufficiently large to avoid this could happen under any optimal policy.  The action space U can then be defined as 

\begin{align}
    U := \left\{\left[-u^{\min}(S_t), u^{\max}(S_t)\right]\mid S_t \in S, \text{s.t. } (1) \text{ and } (2), t \in \mathcal{T} \right\}.
\end{align}

The reward $r(u(S_t))$ of taking action $u(S_t)$ is the sum of the revenues and costs during period $t$ as a result of interacting with the grid.  It is defined as
\begin{align}
    r(u(S_t)) = uc_t + \mathbb{I}_{\{u \leq 0\}}c^+ + \mathbb{I}_{\{x_{t+1}\mid u < 0\}}s
\end{align}
where $\mathbb{I}_{(\cdot)}$ equals 1 if $(\cdot)$ evaluates to true, and is 0 otherwise. Here, by slight abuse of notation, we denote with $x_{t+1}\mid u$ the hypothetical inventory level at time $t+1$ given action $u(S_t)$. If that is negative, demand is unmet and penalty costs $s$ are incurred.

Note that if energy is sent to storage, that is for any state $S_t$, $-u +\bar{y}_t \geq 0$, the amount of energy that is stored depends on the conversion efficiency $\alpha$. To avoid numerical issues associated with round-trip conversion losses and resulting fractional numbers, the total round-trip efficiency is only calculated when sending energy to storage. 

\subsection{Discretization}

For the numerical analysis, we need to discretize the state space. We discretize the amount of hydrogen inventory, the net production throughout the year, and the observed prices. We define $\mathcal X$  as the set of possible net inventory levels, where $\Delta x_t$ represents the interval size of the inventory levels. The intervals to which an action $u$ belongs are split into equally-sized intervals which correspond to the discretization of the inventory levels $\Delta x_t$. We denote the discretized set of actions by $\mathcal U$. The stochastic net solar energy production $\bar Y_t$ is discretized according to $\Delta j_t$, and electricity prices are discretized with $\Delta c_t$. Accordingly, 


\begin{align}
    \mathcal X &= \{0, \Delta x_t, 2\Delta x_t, \ldots, m\}\\
    \bar{\mathcal Y_t} &= \{l^-_t, \Delta j_t, 2\Delta j_t, \ldots, l^+_t\}\\
    \mathcal C &= \{0, \Delta, 2\Delta c_t, \ldots, C\}.
\end{align}

\subsection{MDP for storing, buying from or selling to the grid}

The selling and buying policies result in an inventory process over time in which an immediate reward of $r(u(S_t)))$ is earned after choosing an action at the end of period $t$. The action is chosen after the solar production and electricity prices have been fully observed at the end of a period. Therefore, the decision-maker knows with certainty to which new inventory level the action will lead in the next period. The future inventory $x_{n-1}$ in period $n-1$ with $n$ periods to go $n= T - t$ can be defined as\footnote{For readability, we ignore the case that inventory becomes negative, but this is trivially excluded by taking the maximum of $x_{n-1}$ and 0}

\begin{align}
  x_{n-1}(u, \bar{y}_n)  = x_n + 
  \begin{cases}
  \alpha(u + \bar{y}_n) & \text{if } u+\bar{y}_n \geq 0\\
  u + \bar{y}_n & \text{if } u + \bar{y}_n < 0 
  \end{cases}
\end{align}

 The transition probability~$p_{n-1}(\bar{y}_{n-1}, c_{n-1}, c_{n})$ is defined as the probability of net production realization of $\bar{y}_{n-1}$ and a price realization of~$c_{n-1}$ in period~$n-1$ given price $c_{n}$ in period $n$. Similar to \citet{Zhou2019}, we assume that the underlying stochastic processes are independent. 
We define $V_0(S_0)$ as the total expected profit at the end of the horizon with $n = 0$ periods to go. We assume
\begin{align}
  V_0(S_0)  = 0
\end{align}

For all other time periods in which $n > 0$, action $u(S_n) \in \mathcal U$ can be executed. Accordingly, we define 

\begin{align}
    V_n(S_n)  = \max_{u(S_n) \in \mathcal U} \left \{ r(u(S_n)) + \sum_{\bar{y}_n \in \bar{\mathcal{Y}_n}} \sum_{c_n \in \mathcal{C}}p_{n-1}(\bar{y}_{n-1}, c_{n-1}, c_{n}) V_{n-1}(S_{n-1}) \right \} \label{eq:bel}
\end{align}

Via backwards dynamic programming, $V_n(S_n)$ can be obtained for all periods-to-go $n \in \mathcal{T}$. The associated optimal periodic policy $(u^*_1(S_1), u^*_2(S_2), \ldots, u^*_T(S_T))$ that minimizes long-term average rewards is then obtained by iteratively applying backwards dynamic programming upon this system, where $V_0(S_0)$ is calculated according to  equation $\eqref{eq:bel}$ with $V_{n-1}(S_{n-1})$ equal to $V_{T}(S_T)$ of the previous iteration (equaling zero for the first iteration). Convergence to optimality is proven if all values $V_n(S_n)$ change with the same value (i.e., the long-run average reward) between iterations \citep{puterman2014markov}.

\section{Numerical Analysis}
\label{sec:numeric}

We start our numerical section by introducing a base-case system for which we provide a detailed numerical analysis (see Section~\ref{sec:basecase}) and then describe how the price and production processes are fitted (see Section~\ref{sec:fitdata}). We end the section by examining the optimal policy for the base-case system while focusing on the differences between summer and winter. To provide a comparison for the viability of the base-case system, we compare the optimal policy to a system without any hydrogen storage options. A more extensive sensitivity analysis in which all system parameters deviate one-by-one is postponed to Section~\ref{sec:sensitivity}.

\subsection{Base-case system}
\label{sec:basecase}

The experiments are based on a planned project of a rural village in the Netherlands in which electricity needs are supplied by a solar park. It comprises a hypothetical solar park with a peak capacity of 5 MWp that is connected to a local electricity grid with a maximum capacity of 1.25 MW (which corresponds to 30 MWh per day). Distribution capacities of 2.5 MW are common in practice for lines that operate at the distribution (local) level rather than the (national) transmission level. In our experiments, we set a more constrained distribution capacity of 1.25 MW, to represent the situation in which the distribution capacity is more constrained. We assume the solar park is connected to a 2.1 MW electrolyzer and a 2.1 MW stack of fuel cells. For both of the electrolyzer and fuel cell, this translates to a capacity of 50 MWh per day. Since only relatively small amounts (up to 1 MWh) can be stored inside pressurized vessels, we assume that hydrogen is stored into a co-located large-scale storage location with a capacity of 1000 MWh. Moreover, we assume a round-trip efficiency~$\alpha$ of 0.5. 

\subsection{Fitting the price, production, and demand process}
\label{sec:fitdata}

Day-ahead hourly wholesale electricity prices in euro/MWh in the Netherlands between 2015 and 2019 are obtained from ENTSOE Transparency Platform \citep{entsoe}. These prices are aggregated to daily prices using the intra-day mean. We assume that the electricity prices which apply to the storage owner exhibit similar behavior to wholesale day-ahead electricity prices. 

The average daily day-ahead prices exhibited strong autocorrelation (0.872 for a time lag of 1). Moreover, weekly autocorrelation was observed in which autocorrelation was stronger for weekly time intervals than for intra-week intervals. For example, the autocorrelation decreases to 0.773 for lags up to 6 and jumps to 0.815 for lag 7, suggesting that weekday effects are existent. Seasonal effects are not directly apparent from the data. 

Inspired by existing approaches in literature (e.g., \cite{Zhou2019}),  
we test three different AR(1) models of the form $C_t =\phi + \theta C_{t-1} + \xi_t$, where $\xi_t \sim N(0, \sigma^c)$, $C_t$ is the predicted value, $C_{t-1}$ is the observation at $t-1$, $\phi$ is a constant, and $\theta$ is the AR term with a time lag of 1. First, we fit an AR(1) process to the daily electricity prices in which both monthly and weekday effects are removed from the original observations. Secondly, we fit an AR(1) process to prices in which only weekday effects are removed. Thirdly, we fit an AR(1) process to the original observations. We compare the models by evaluating the standard error of the estimate in relation to the actual observations. The month day and weekday effects are removed similarly as in \cite{Zhou2019}. To evaluate the fit of the AR(1) models, the standard error was calculated as $\sqrt{\sum_{t=2}^{T}(C_{t} - \hat{C}_t)^2/T}$, where $T$ is the number of periods and $\hat{C}_t = \phi + \theta C_{t-1} + f'(t) + \xi_t$ is the predicted price in period $t$. The seasonality effects of the electricity prices are described by $f^{'}(t)$ which is defined similarly as in \cite{Zhou2019}. Because the standard error of the model which was fitted to the original data is lowest (7.7), as is given in Table~\ref{table:AR_errors}, we do not include month and weekday effects in our AR(1) process.


 
\begin{table}[ht]
\small
    \centering
	\caption{AR(1) parameters and standard error}
	
	\begin{tabular}{llll}
		\toprule
		 Model & $\pi$ & $\phi$ &Std. error \\
		\midrule
           1. Remove both month and weekday effects & -0.004 & 0.89 & 37.7\\
           2. Remove weekday effects & -0.006 & 0.91 & 38.3\\
           3. Fit to original data & 5.23 & 0.87 & 7.7\\
		\bottomrule
	\end{tabular}
	\label{table:AR_errors}
\end{table}

To model the demand in our base-case system, we assume that the solar park and hydrogen fuel cells connect directly to a set of 1500 houses that are responsible for the local electricity demand. We have obtained data on electricity consumption from the society of the Dutch Energy Data Exchange (NEDU). The data represents average electricity consumption levels per 15 minutes as a fraction of the total yearly consumption level for 3001 measurements during the years 2016, 2017, and 2018. Since the data is highly aggregated, the data can be scaled to 1500 households to represent our base-case system. We assume that one household on average consumes 2990 kWh in electricity per year \citep{nibud_2019}. The scaled daily consumption levels, as used in our base-case system, have a minimum of 9.9 MWh, a mean of 12.3 MWh, and a maximum of 16.1 MWh per day. 

The data exhibits a strong linear relationship with time in which marginal consumption levels are negative between winter and summer and positive between summer and winter. Accordingly, we split the data into two subsets of observations and fit a linear regression to each subset. The splitting procedure was based on minimizing the sum of the standard errors for both models. Accordingly, model 1 was based on day 1 to 199, whereas model 2 was based on day 200 to 365. The fitted models are displayed in Table~\ref{table:lm_consumption}. Since the original data is highly aggregated, we assume the data is normally distributed (i.i.d.) with a daily $\mu_t$ and a constant standard deviation $\sigma$. Since the standard errors are relatively similar, we chose the highest standard error of both models ($\sigma=0.62$) as the standard deviation of electricity demand in the experiments.

\begin{table}[h]
\small
    \centering
	\caption{Linear models on electricity consumption}
	\label{table:lm_consumption}
	
	\begin{tabular}{llll}
		\toprule
		 Model &  Intercept & Estimate & $\sigma$ \\
		\midrule
           Day 1 to 199 & 15.3& -0.0302& 0.62\\
           Day 200 to 365 & 1.79& 0.0372& 0.55\\
		\bottomrule
	\end{tabular}

\end{table}

We assume that the daily solar energy production levels are stochastic. For each day, hourly solar energy production levels between 2005 and 2016 were obtained from PVGIS \citep{pvgis_2016} and aggregated to daily amounts. The production levels correspond to an installed capacity of 5 MWp. The original data was aggregated to daily production levels. For each week, the daily observations within the week of 11 years were normalized to a range between 0 and 1 and fitted to a beta distribution to increase the number of observations. Accordingly, daily solar energy production levels are represented by shape parameters~$\alpha_t$ and~$\beta_t$. Beta distributions are commonly used in modeling daily solar energy production levels \cite{YOUCEFETTOUMI200247, KOUDOURIS2017398}. Similar to \citet{Boland2020} and \citet{soubdhan2009classification}, we assume that~$Y_t$ is independent and identically distributed for each period. 

\subsection{Optimal Policy Structure}
\begin{figure}[t!]
	\centering

	\resizebox{380pt}{!}{%
		\includegraphics[]{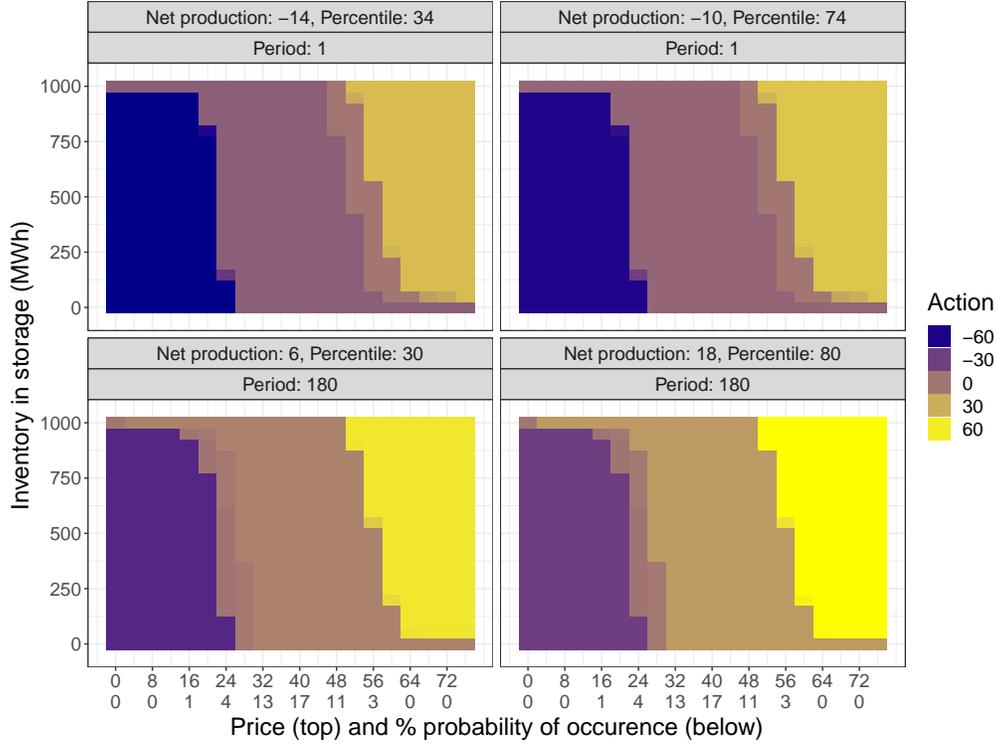}}
	\caption{Optimal policies for winter (top) and summer (below) and for the first (left) and third (right) quantiles of net production}
		\label{figure:optimalPolicy}
\end{figure}

We solved the MDP associated with the base-case system employing backward dynamic programming. As our states are dependent on the day of the year, we considered 50 years. The results presented correspond to the optimal policy of year 0, and can be interpreted as the policy that maximizes long term average rewards (see, for a similar approach, \citet{byon2010season}). Implementation is done in $C$++$17$ and the MDP is solved on an Intel Xeon 2.5Ghz processor using 4 threads. In the following, we discuss the structure of the optimal policy of the base-case system. The performance of this policy is discussed separately in Section~\ref{sec:policyPerformance}.

Figure~\ref{figure:optimalPolicy} shows the optimal policies (i.e., the amount of electricity to sell to the grid) for a period in the winter (Period 1) and in the summer (Period 2). For both periods, we present the optimal policies for the first and third quantile of the net production distribution. On the x-axis, we show the observed electricity price and the probability of occurrence. On the y-axis, we portray the inventory level of the hydrogen storage. In this way, the four graphs represent a cloudy winter day (top left), a sunny winter day (top right), a cloudy summer day (bottom left), and a sunny summer day (bottom right).

From Figure~\ref{figure:optimalPolicy}, one can observe four different types of actions that are taken in the optimal policy, in any of the depicted situations: First, if prices are relatively low the optimal policy prescribes to buy as much electricity from the grid as possible. Second, if prices start to increase, it is best to buy or sell the observed net production. Third, dependent on the actual day of the year, it might be optimal to not buy or sell electricity from the grid, as long as enough inventory is on hand. Fourth, if prices are high enough, it is best to sell as much electricity as possible. We observe that the action to not interact with the grid takes place for opposite inventory levels in the summer and the winter. In the summer, ``no interaction" is optimal for relatively high price levels and the net production is then converted to hydrogen, in the winter net shortages are fulfilled by converting hydrogen to electricity, which also avoids interaction with the grid.

\begin{figure}[ht!]
	\centering

	\resizebox{350pt}{!}{\includegraphics[]{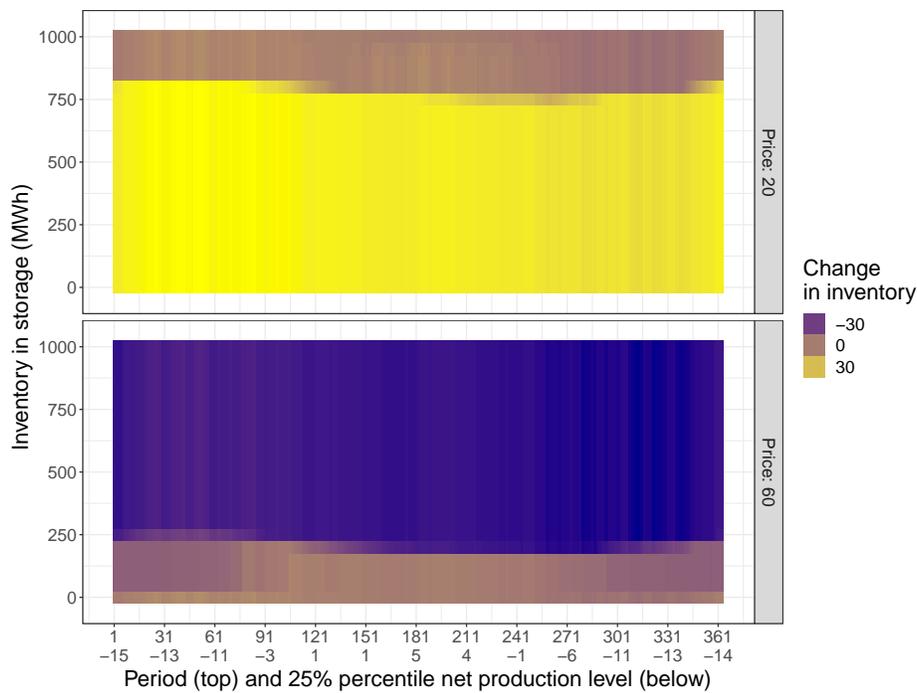}}
		
	\caption{Optimal policies as the change in inventory ($\Delta x_t$) for a low price ($c_t=20$, top) and high price ($c_t=60$, below), and a net production percentile of 25\%}
	\label{figure:optimalPolicy2}
\end{figure}

\begin{figure}[ht!]
	\centering

	\resizebox{350pt}{!}{\includegraphics[]{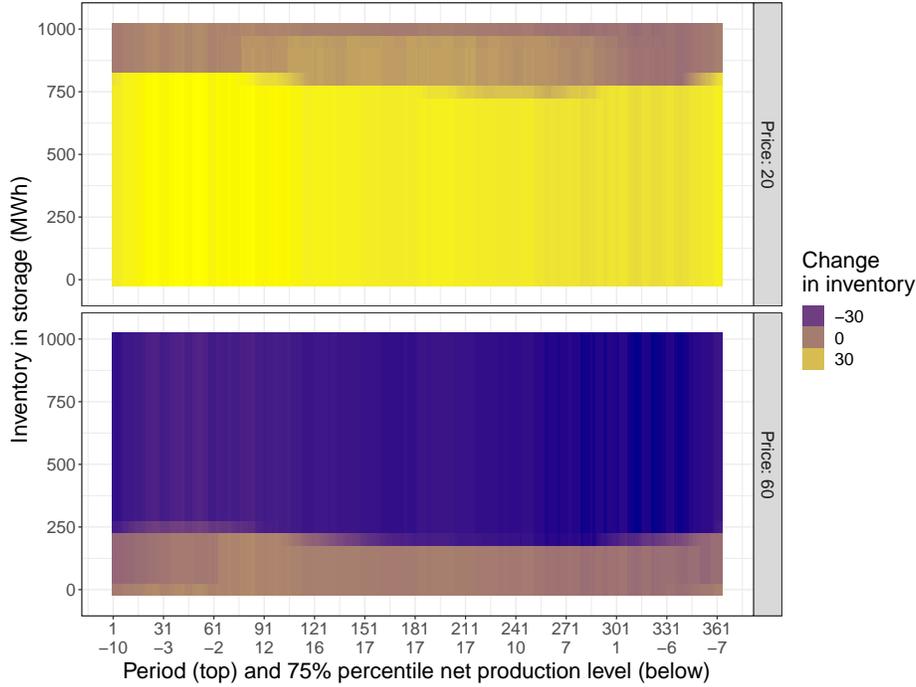}}
		
	\caption{Optimal policies as the change in inventory ($\Delta x_t$) for a low price ($c_t=20$, top) and high price ($c_t=60$, below), and a net production percentile of 75\%}
	\label{figure:optimalPolicy3}
\end{figure}

To better understand how the optimal policy differs throughout the year, Figure~\ref{figure:optimalPolicy2} presents the optimal action as a function of time and the inventory level for the 25\% net production percentile and for the two price levels ($c_t=20$ and $60$). Figure~\ref{figure:optimalPolicy3} does the same for the 75\% net production percentile. The action is represented as the resulting change in inventory level. 

It is observed that the change in inventory has a period-dependent threshold. In the summer (the middle part of both pictures), the inventory at which optimal actions lead to an inventory increase is lower than in the winter, due to oversupply of electricity in the summer and shortages in winter. The inventory levels at which these thresholds occur are similar for both low and high net production levels. 
Figure~\ref{figure:optimalPolicy3} shows that inventory-increasing actions are also prevalent in summer when prices are low (i.e., $c_t = 20$). This indicates the prevalence of seasonal effects in the optimal policies.

This behavior can be attributed to the following dynamics. Early in the year, a higher probability exists of encountering future electricity shortages than later in the year. Therefore, energy is stored at times of low prices early in the year to enable accumulating sufficient inventory for moments of shortages later in the year. Furthermore, buying decisions made early in the year facilitate the potential to benefit from price differences later in the year. These dynamics will be explained in more detail in Section~\ref{sec:policyPerformance}.

%

\subsection{Optimal Policy Performance}
\label{sec:policyPerformance}

We simulated the optimal policy of the base-case system for a total of 1,100,000 years, using the first 100,000 years as a warm-up for the simulation. Key performance indicators are given in Table~\ref{table:policy_KPI}. As a benchmark, we also provide the statistics of our base-case system if no hydrogen storage is available (BM1), and in case the optimal policy for stationary net production (i.e., ignoring seasonal effects) is applied to our base-case system (BM2).
Our key performance indicators are mean profits, electrolyzer utilization, and the mean percentage of the time in which congestion occurs at the cable connected to the solar park. Similar to \cite{creti2019}, we define congestion as the event in which the amount of electricity sent to or obtained from the distribution grid equals the distribution capacity to which the supplier or consumer, in this case, the solar park with storage, is connected. Accordingly, the mean percentage of time congestion is measured as the mean percentage of time in which selling or buying energy equals the grid distribution capacity.
The electrolyzer utilization is given as the percentage use of its full capacity. 
\label{sec:policuPerformance}
\begin{table}[ht!]
\small
    \centering
	\caption{Summary statistics reference case}
	
	\begin{tabular}{lrrr} 
		\toprule
		   KPI & Base-case system & BM1 (no storage) & BM2 (ignoring seasonality) \\ \midrule
           Mean profit per year & 6579.4 & -4060.5 & 6387.0\\
           Mean electrolyzer utilization (\%) &22 & - & 22.4\\
           Mean \% time congestion (\%) & 8.6 & - & 8.8\\
		\bottomrule
	\end{tabular}
	\label{table:policy_KPI}
\end{table}

From Table~\ref{table:policy_KPI}, we observe that adding storage increases mean profit per year by from -4060.5 to 6579.4. It is also clear that ignoring seasonality is suboptimal, as the mean profit per year decreases by 2.9\% comparing BM2 to the base-case system. The mean electrolyzer utilization denotes the amount of electricity converted to hydrogen given that the electrolyzer is used. It increases 1.6\% when seasonality ignored. The mean percentage of time the cable is used to its full capacity, reflecting situations in which congestion occurs at the cable to which the solar park is connected, equals 8.6\% and 8.8\% for the base case and when ignoring seasonality (BM2), respectively. 

\begin{figure}[ht!]
	\centering

	\resizebox{350pt}{!}{\includegraphics[]{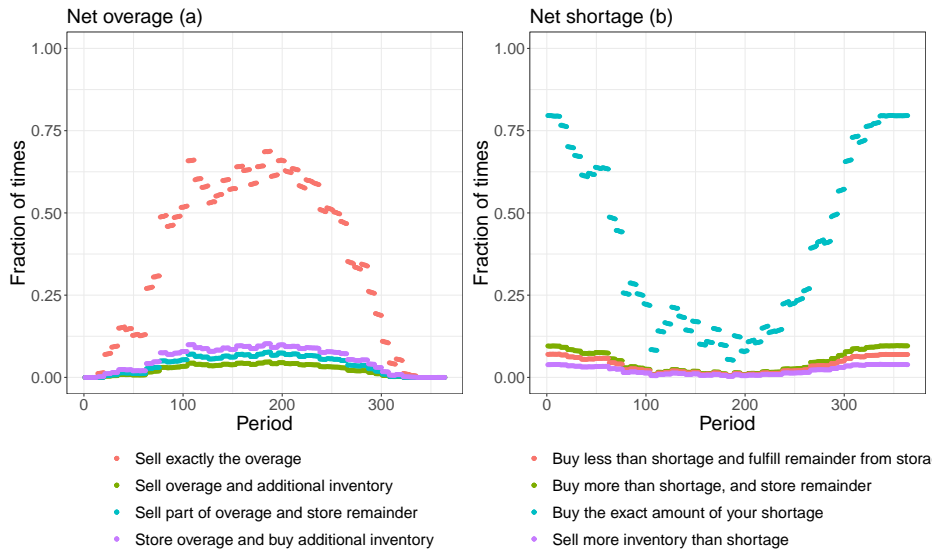}}
	
	\caption{Fraction of time a certain action occurs over time during a net overage (left) or shortage (right)}
	\label{figure:actions_scenarios}
\end{figure}

We further detail the expected number of times particular actions are taken throughout the year. In Figure~\ref{figure:actions_scenarios}, we detail these actions for a net overage (a) and a net shortage (b).

Given a net shortage, the red points indicate the fraction of times less than the shortage is bought while the remainder is obtained from storage. The green points indicate that more is bought than the shortage while the remainder is stored. The blue points indicate the fraction of times the exact amount of the shortage is bought. The purple points indicate that more inventory is sold than the shortage.  

Given a net overage, the red points indicate the fraction of times exactly the net overage is sold. The green points indicate the fraction of times the overage is sold plus additional inventory. The blue points indicate the part of the overage that is sold while the remainder is stored. The purple points indicate the fraction of times the overage is stored and additional inventory is bought. 

In the case of a net overage for the base-case system, the red points in Figure~\ref{figure:actions_scenarios} (a) show that policies in which exactly the overages are sold to the grid are most prevalent. These follow the seasonality pattern associated with solar electricity production and occur most frequently at~69\% of the time in the summer. Policies in which additional electricity is sold out of storage are associated with exploiting price difference possibilities. These are the least common during overages and are highest in summer occurring~5\% of the time. Selling less than the overage is also not a common strategy and occurs maximally at~7.7\% of the time. This indicates that storage is not used frequently in cases of overages. This also suggests that excess net production can at best be sold directly to the grid to avoid conversion losses when using storage, even at low prices.

For cases of net shortage in Figure~\ref{figure:actions_scenarios} (b), occurrences in which the exact shortage is bought from the grid are most prevalent. The blue points show that the fraction of times the exact amount of the shortage is bought follows an inverse pattern compared to policies in which exactly overages are sold during net overages. These are highest in winter up to~79\% and lowest in summer down to~5\%. Other policies are very uncommon for the conducted experiments.

Even though these policies are relatively uncommon, the less frequent action types that include not simply selling or buying net production differences, are important for the feasibility of a solar park in rural, possibly congested, areas. For instance, in The Netherlands, it is not allowed to install a solar park with a maximum capacity higher than the distribution capacity, even peaks only occur at clear summer days. The results presented in Figure~\ref{figure:actions_scenarios} indicate that using a storage facility will not interact structurally different from a classic solar park without storage, only its distribution capacity is limited. These are exactly the moments when the less-frequent actions depicted in Figure~\ref{figure:actions_scenarios} play an important role to keep the base-case system feasible. This includes the ability to exploit price differences and to fulfill shortages at times when prices are high. As can be seen in Table~\ref{table:policy_KPI}, a solar park without a storage facility (connected to local electricity demand) yields negative mean profits when demand needs to be solely fulfilled from the electricity grid.  

\section{Sensitivity Analysis}
\label{sec:sensitivity}

We further investigate the performance of our system, using the key performance indicators already presented in Section~\ref{table:policy_KPI}. We first investigate the impact of changing the distribution capacity (Section~\ref{sec:sens_cable}) and afterward discuss the impact of the storage capacity on the performance of the system (Section~\ref{sec:sens_storage}). For the distribution and storage capacity, we also provide insights into the interaction with the grid, as this is relevant for future solar park owners and legislators due to its relation to grid congestion issues. We end this section by concisely showing the impact of changing the electrolyzer capacity, conversion efficiency, price mark-up, and production capacity in Sections~\ref{sec:sens_elec}-\ref{sec:sens_prodCap}. In each section, we only vary the parameter that is being discussed and set the other parameter values equal to the base-case system.

\subsection{Distribution capacity}
\label{sec:sens_cable}

\begin{figure}[ht!]
	\centering

	\resizebox{400pt}{!}{\includegraphics[]{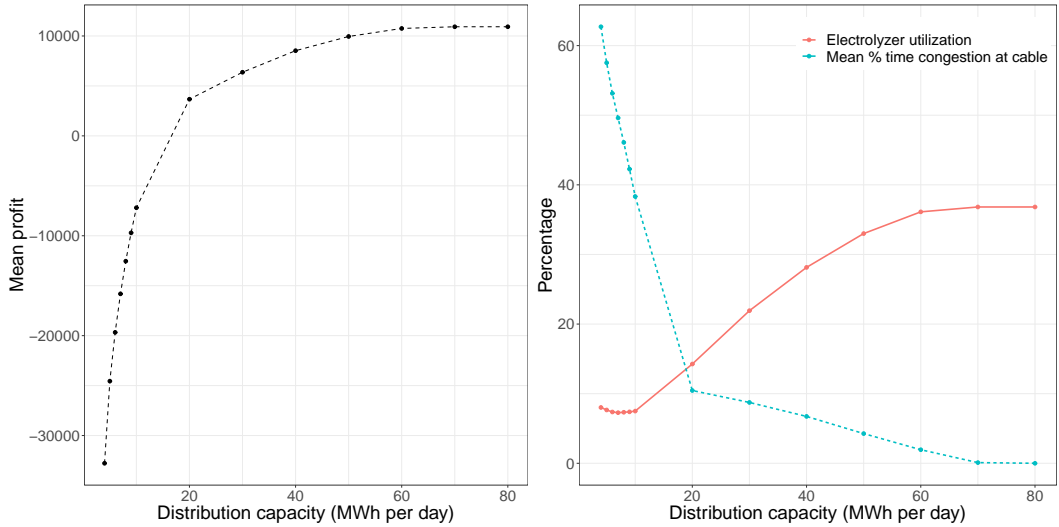}}
	
	\caption{Summary statistics and distribution capacity ($k^c$)}
	\label{fig:sensitivity_distribution}
\end{figure}

We vary the distribution capacity between 1~MWh to 80~MWh per day, which corresponds to~0.04 to~3.3 MW. 
In Figure~\ref{fig:sensitivity_distribution}, we see that the mean profit increases for larger distribution capacities, between 4~MWh and 80~MWh per day, because our system becomes less constrained. Distribution capacities below 4~MWh per day are infeasible for our parameter settings, due to unmet demand. 

Low distribution capacities up to 10 MWh per day lead to negative profits, which is due to the limited possibility to exploit price differences as local demand should always be satisfied first. For increasing distribution capacities, the electrolyzer utilization increases up to~36.8\%. This due to the exploitation of price differences. If prices are low, electricity is bought from the grid to sell it again when prices are high. Finally, the percentage of the time the distribution capacity is fully utilized with congestion at the connected cable approaches~0\%, which is expected since the distribution capacity is then only constraining the system for high net production overages.

\begin{figure}[ht!]
	\centering

	\resizebox{400pt}{!}{\includegraphics[]{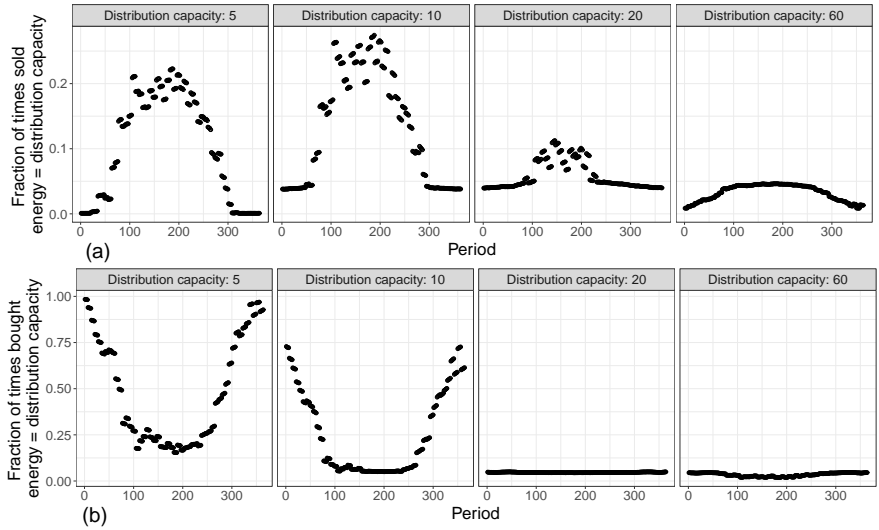}}
	
	\caption{Fraction of time that sold (a) and bought (b) energy equals the distribution capacity (selling and buying-induced congestion)}
	\label{figure:distr_cap}
\end{figure}

Figure~\ref{figure:distr_cap} (b) (bottom) shows the fraction of times in which the amount of energy bought equals the distribution capacity for different levels of distribution capacity. We label this event as buying-induced congestion that takes place at the connection with the electricity grid. Figure~\ref{figure:distr_cap} (a) (top) shows the fraction of times in which sold energy equals the distribution capacity, causing selling-induced congestion.

Figure~\ref{figure:distr_cap} shows that relatively low levels of distribution capacity (e.g., $k^c = 5$) cause a combination of buying-induced and selling-induced congestion. This is the result of preventing future shortages and exploiting price difference opportunities. Both types of congestion follow a seasonal pattern. Whereas buying-induced congestion is highest in the winter months, selling-induced congestion is highest in the summer. This can be attributed to net production shortages that occur in the winter and net production overages that occur in the summer. In line with the results in Figure~\ref{figure:actions_scenarios}, this indicates that selling net overages and buying net shortages from the grid is a preferred action in general.

When the distribution capacity becomes larger (i.e., $k^c = 10$), more selling-induced congestion occurs since the optimal policy is less impacted by possible shortages in winter. Consequently, less energy is stored and more energy is sold to exploit price differences. Buying-induced congestion simply decreases for higher distribution capacities as the distribution capacity becomes less-often the limiting factor when net-shortages occur. When capacity increases even more (i.e., $k^c = 20, 60)$, the occurrence of both types of congestion decreases, as the net-production realizations can be sold or bought completely from the grid without being restricted by the distribution capacity. Price differences are exploited in higher quantities, while fewer congestion events are observed.

\subsection{Storage capacity}
\label{sec:sens_storage}

\begin{figure}[h!]
	\centering

	\resizebox{400pt}{!}{%
		\includegraphics[]{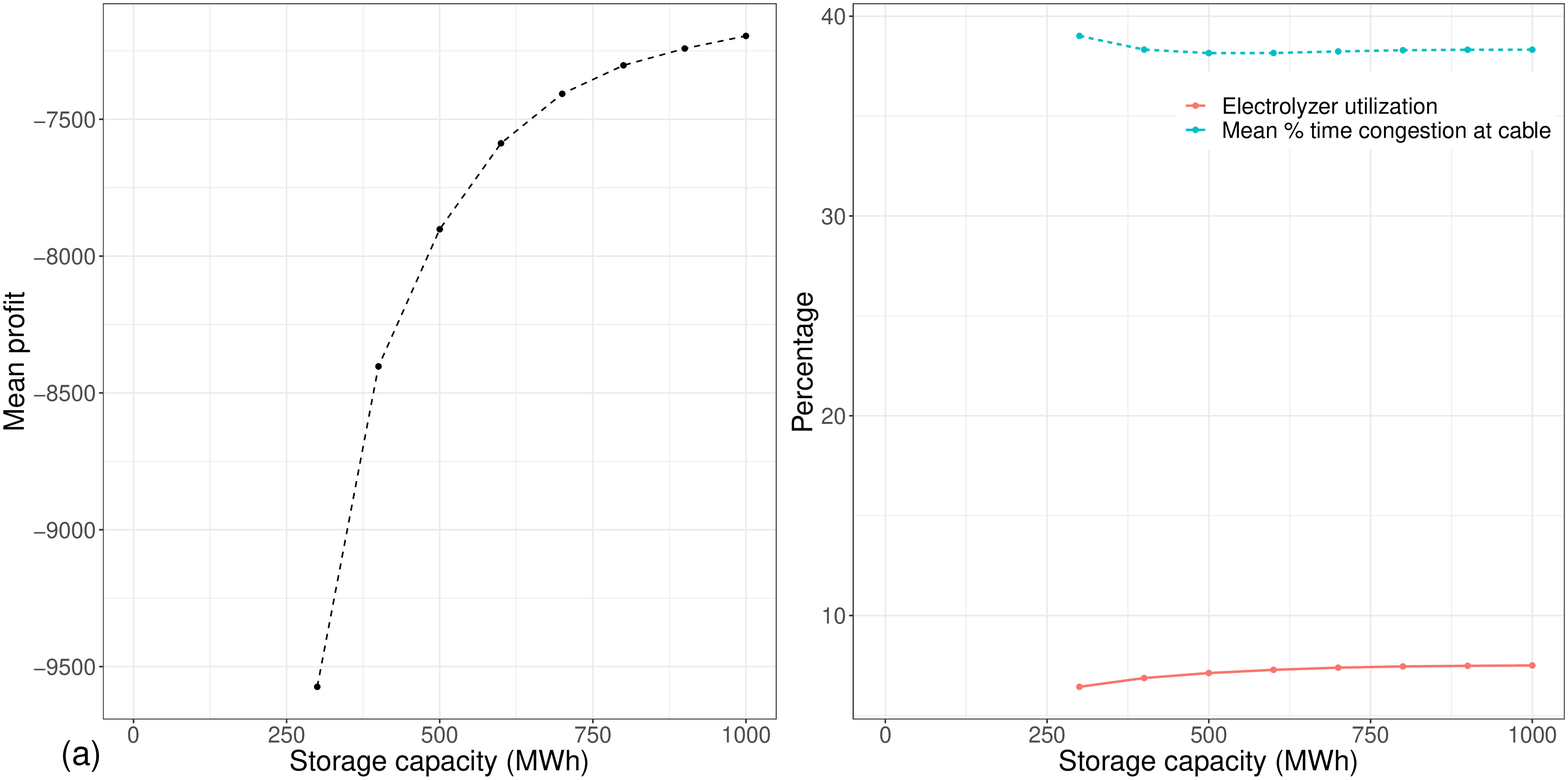}}
	\resizebox{400pt}{!}{%
		\includegraphics[]{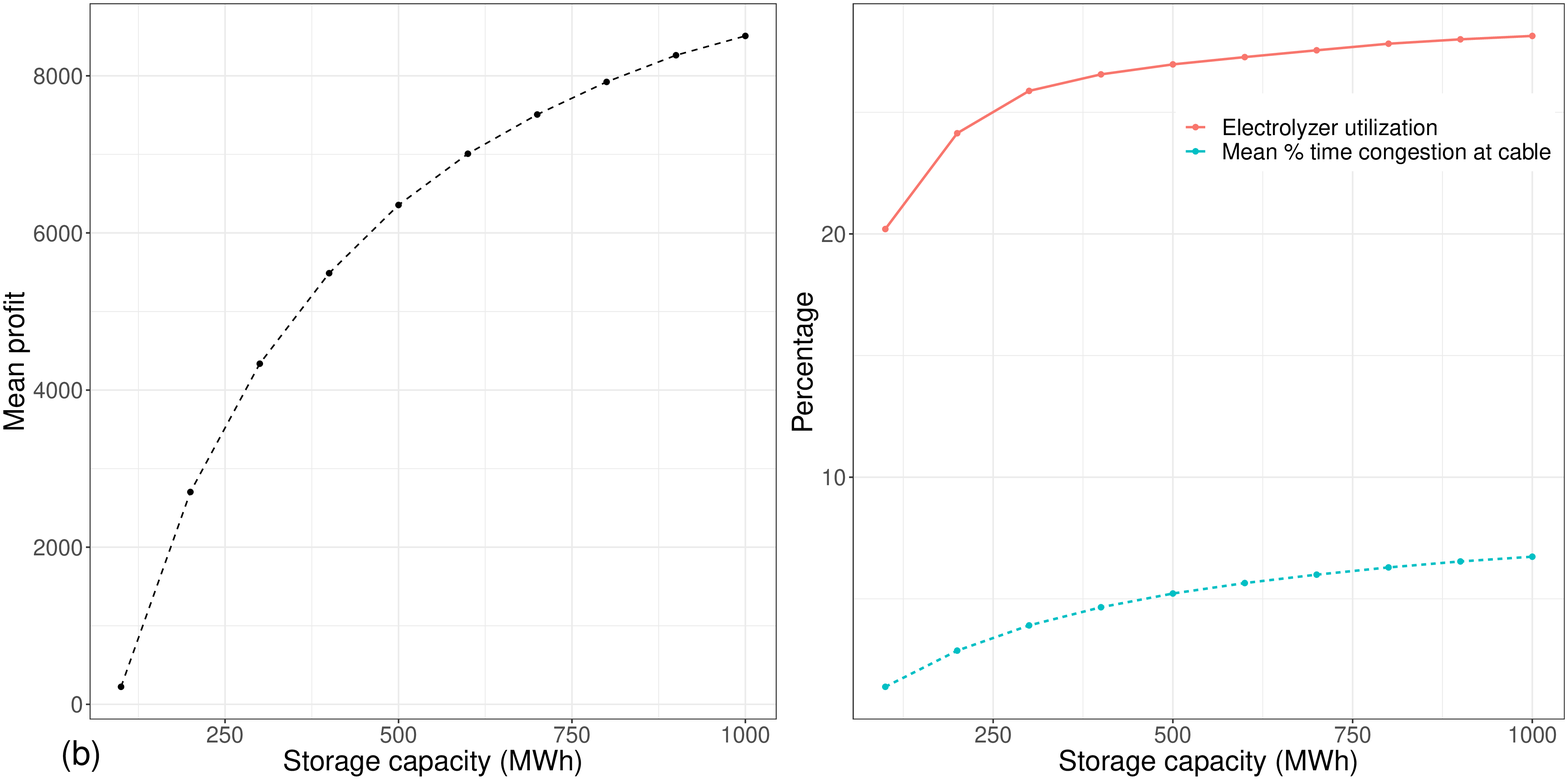}}
	
	\resizebox{400pt}{!}{%
		\includegraphics[]{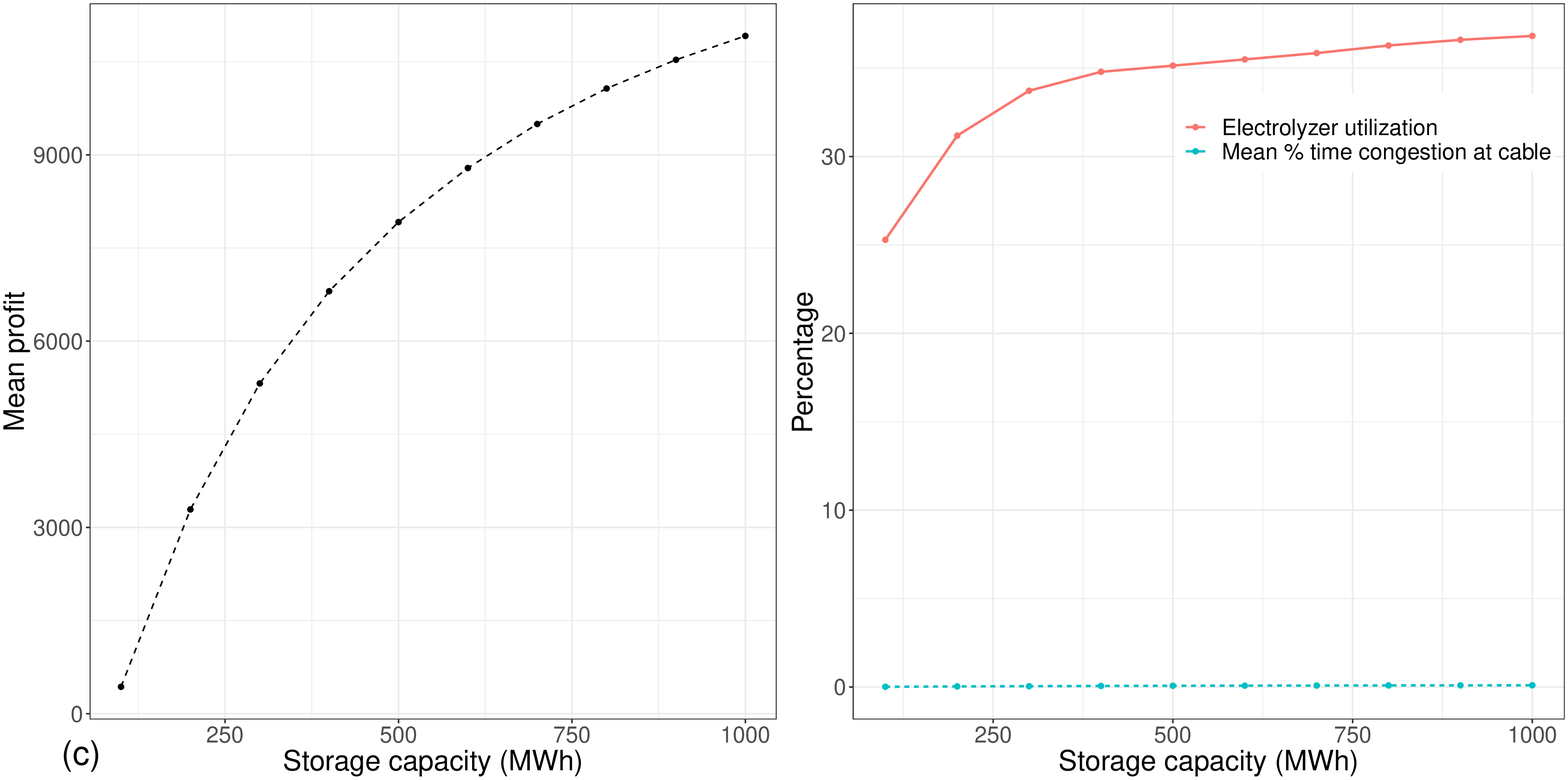}}

	\caption{Summary statistics for varying storage capacity, for distribution capacities equal to $k^c=10$ (a), $k^c = 40$ (b) and $k^c = 80$ (c)}
		
	\label{figure:storage_k10}
\end{figure}

While the base case used a storage capacity of 1000 MWh and a distribution capacity of 1.25 MW (30 MWh per day), we vary the storage capacity between~100 and 1000~MWh with increments of~100, for three different levels of distribution capacity (10, 40 and 80 MWh per day), see Figure~\ref{figure:storage_k10}. In this way, we investigate how storage can facilitate congestion reduction when distribution capacity is constrained. In addition, it allows us to study how profits are affected when distribution capacity is sufficient.

Figure~\ref{figure:storage_k10} shows negative mean profits which increase at a marginally decreasing rate with storage capacity for each distribution capacity. For distribution capacity $k^c = 10$ (a), the mean profits are negative due to the highly constrained distribution capacity.  Note that storage capacities smaller than 300~MWh are not displayed in Figure~\ref{figure:storage_k10} (a). This is because these capacities result in systems where it is not possible to always satisfy local demand. Here, the percentage of unmet demand ranges between 0.005\% and 8\% and a penalty for unmet demand is incurred.  Electrolyzer utilization levels remain relatively constant (between~6.9\% and~7.5\%), and congestion levels are not affected by the storage capacity of 300~MWh or higher (the mean percentage time congestion remains between~38.2\% and~38.3\%). These results indicate that the distribution capacity is too limited across all levels of storage capacities to enable positive profit levels, more effective use of the electrolyzer and to reduce congestion issues. 

Results for $k^c = 40$ and $k^c = 80$ in Figure~\ref{figure:storage_k10} (b) and (c) show that congestion issues are not relevant anymore for our distribution capacity. Furthermore, increasing storage capacity to $k^c = 40$ leads to more congestion at the cable connection, due to increasing interactions with the electricity grid. This is in line with the observed electrolyzer utilization for higher storage capacities. 

Concluding, hydrogen storage used to supply electricity does not lead to profits when distribution capacity is too small and the storage owner aims to maximize profits. Moreover, seasonal storage also does not aid in reducing local congestion at the connected cable when the associated distribution capacity is too small. We advise to carefully decide upon the offered distribution capacity and the installed storage capacity, as this impacts the frequency at which distribution capacity is fully utilized, and, therefore, affects local congestion issues. 

\subsection{Electrolyzer capacity}
\label{sec:sens_elec}

\begin{figure}[ht!]
	\centering

	\resizebox{400pt}{!}{%
		\includegraphics[]{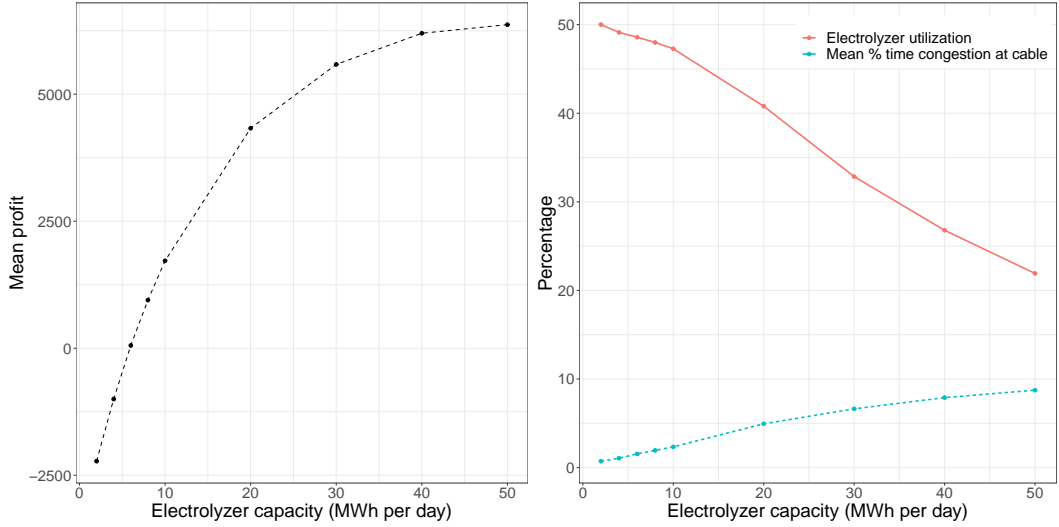}}
	
	\caption{Summary statistics and electrolyzer capacity ($k^+$)}
	\label{figure:electrolyzer}
\end{figure}

We vary electrolyzer capacities $k^+$ between 2 and 50 MWh per day. Figure~\ref{figure:electrolyzer} shows that mean profits increase, at a marginally decreasing rate, for increasing electrolyzer capacity. Utilization levels decrease and range between~50\% and~29.9\% for $k^+ = 2$ to $k^+ = 50$. Profits are positive when the electrolyzer capacity is at least 6 MWh per day (250 kW). Without considering capital expenditures, this suggests that over dimensioning the electrolyzer capacity leads to increased profits, even though utilization levels are reduced. These results suggest that electrolyzer utilization is not a good proxy for profitability as large electrolyzer capacities with a relatively low electrolyzer utilization may lead to more profits than smaller electrolyzer capacities with a relatively high electrolyzer utilization. This only applies to operational profits as capital expenditures are not taken into account.

The mean time in which congestion occurs due to the full utilization of the connected cable ranges between~0.7 and~8.7\% for these electrolyzer capacities. This is in line with what is expected, as lower electrolyzer capacities lead to a more constrained system. 

\begin{figure}[ht!]
	\centering

	\resizebox{400pt}{!}{%
		\includegraphics[]{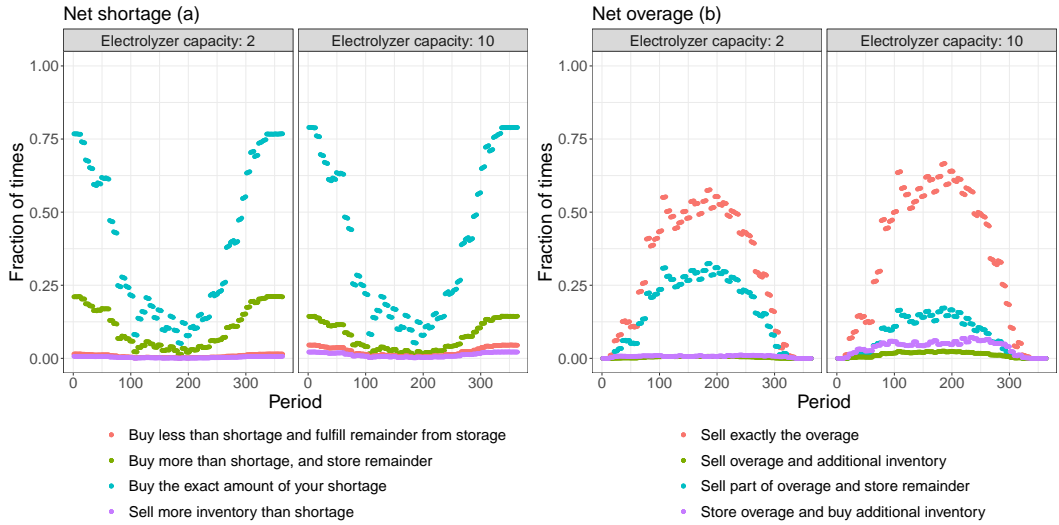}}
	
	\caption{Fraction of time a certain action occurs over time during a net overage (left) or shortage (right)}
	\label{figure:electrolyzer_decisions}
\end{figure}

A change in the type of (optimal) actions taken is observed for different electrolyzer capacities. We portray the two combinations of electrolyzer capacities in Figure~\ref{figure:electrolyzer_decisions} ($k^+ = 2, k^+=10)$. It shows the mean fraction of times an action occurs for a net shortage (a) and a net overage (b). The x-axis indicates the period (day) and the y-axis indicates the fraction of time a particular action occurred. The same actions as in Figure~\ref{figure:actions_scenarios} are given. 

Given a net shortage and a low electrolyzer capacity (i.e., $k^+ = 2$), the green points in Figure~\ref{figure:electrolyzer_decisions} (a) show that buying the exact amount of the shortage is most prevalent in winter and least prevalent in summer. Other actions are almost non-existent. For a high electrolyzer capacity (i.e., $k^+ = 10$), buying more electricity than the shortage occurs 8 percent points less frequently on average than for a low electrolyzer capacity (i.e., $k^+ = 2$).

Given a net overage and a low electrolyzer capacity  (i.e., $k^+ = 2$), the blue points in Figure~\ref{figure:electrolyzer_decisions} (b) show that actions in which part of the overage is sold and the remainder stored are more prevalent than for high capacity (i.e., $k^+ = 2$), indicating that storing part of an overage is mostly needed for low electrolyzer capacities to cover potential future shortages.

These results indicate that the higher electrolyzer utilization at low capacity (i.e., $k^+ = 2$) is caused by buying and storing additional electricity from the grid in times of shortages or storing in times of overages. The stored energy can be used to cover potential future shortages during times of high prices. When capacity is higher (i.e., $k^+ = 10$), the risk of supplying future potential shortages from the grid at high prices is reduced, and storing energy is not beneficial.

\subsection{Conversion efficiency}

\begin{figure}[ht!]
	\centering

	\resizebox{400pt}{!}{%
		\includegraphics[]{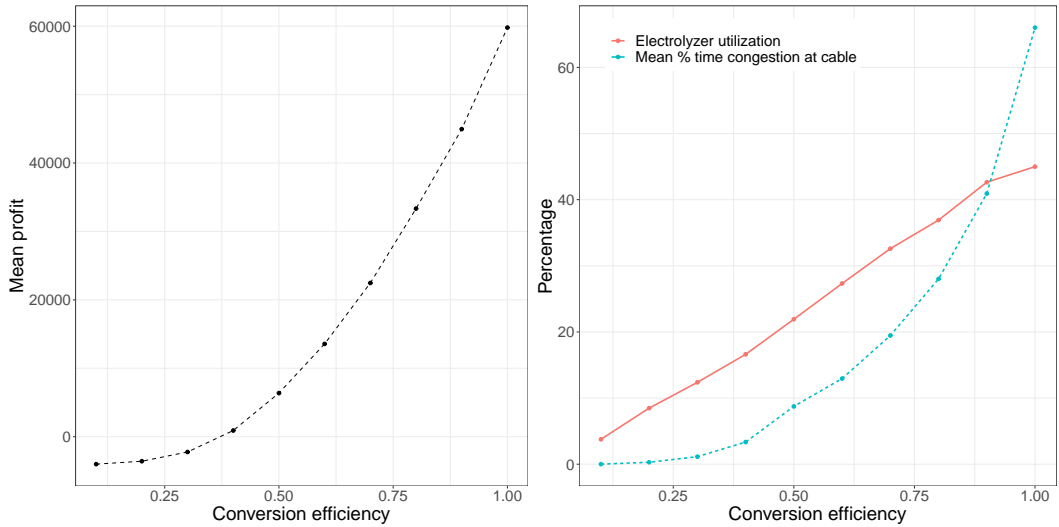}}
	
	\caption{Summary statistics and conversion efficiency ($\alpha$)}
	\label{figure:conversion}
\end{figure}

Figure~\ref{figure:conversion} shows that mean profits are positively related to conversion efficiency since storage becomes increasingly beneficial in both exploiting price differences and covering shortages that do not need to be bought from the grid. For this reason, electrolyzer utilization is also positively related to conversion efficiency. Moreover, congestion at the connected cable increases for higher conversion efficiencies due to increased interaction with the grid. This indicates that technological improvements related to hydrogen storage which lead to higher efficiencies also cause increasing grid interactions. 

\subsection{Price markup}

\begin{figure}[ht!]
	\centering

	\resizebox{400pt}{!}{%
		\includegraphics[]{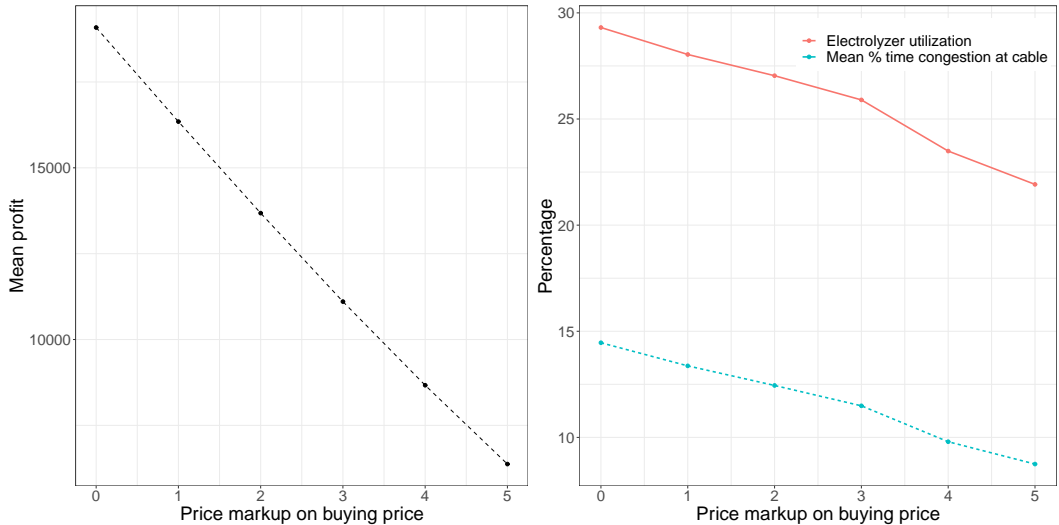}}
	
	\caption{Summary statistics and price markup on buying price ($c^+$)}
	\label{figure:markup}
\end{figure}

We vary the price markup related to buying electricity from the grid (i.e., $c^+$) between 0 and 5. Figure~\ref{figure:markup} illustrates that mean profits are negatively related to the price markup on the buying price. This is expected because price markups on buying electricity discourage the use of storage to benefit from price differences over time. The electrolyzer utilization is reduced from~29\% to~22\% between price markups of 0 and 5. Reduced grid interaction as a result of higher price markups reduces congestion levels as well. This indicates that the electrolyzer is used less to buy energy from the grid for the purpose of benefiting from price differentials enabling the use of storage to prevent congestion.  This confirms that price markups can be used as an instrument to reduce congestion levels in which markups can be raised until using storage is not profitable anymore.  

\subsection{Production capacity}
\label{sec:sens_prodCap}

\begin{figure}[ht!]
	\centering

	\resizebox{400pt}{!}{\includegraphics[]{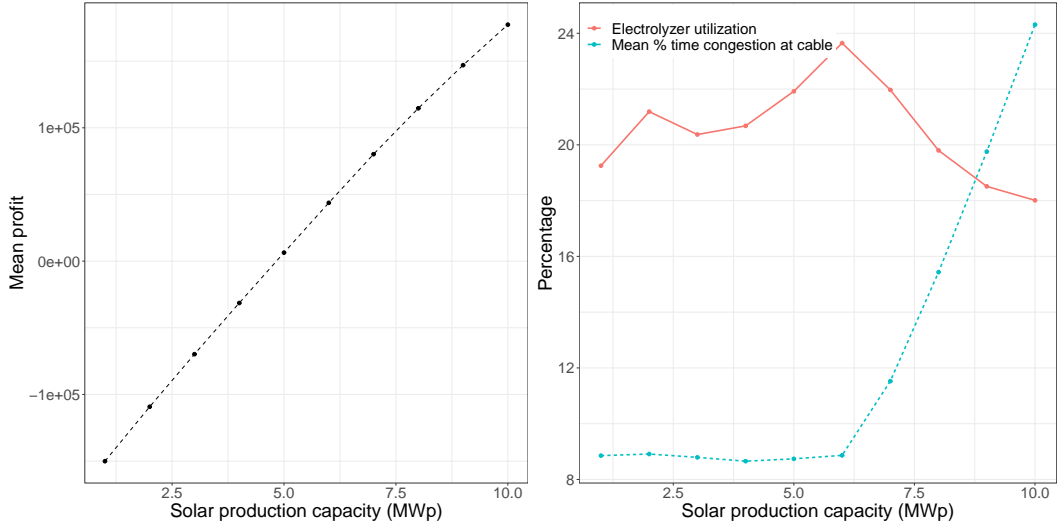}}
	
	\caption{Summary statistics and solar production capacity ($w$)}
	\label{figure:solarcap}
\end{figure}

We vary the production capacities of the solar park (i.e., $w$) between~1 and 10~MWp. Figure~\ref{figure:solarcap} indicates that mean profits appear to increase linearly with increased production capacities. Up to 4~MWp of the conducted experiments, profits are negative, due to a high reliance on the grid to cover shortages and the inability to store energy when prices are low to cover future shortages. Increased reliance on the grid at low production capacities is reflected in the electrolyzer utilization rates, which increase for capacities between 3 and 6~MWp and decrease for higher capacities. At low production capacities (e.g., $w = 1$), the electrolyzer is deployed to store energy which is bought from the grid to cover future shortages. Congestion at the connected cable increases nearly linearly for production capacities above 6 MWp. This is attributed to increased overages which are sold to the grid during summer. 

\section{Conclusion}
\label{section: conclusion}
Increased decentralization of renewable energy sources such as solar parks leads to grid congestion in rural areas where grid distribution capacity is limited. At the same time, supplying local villages in the vicinity of solar parks reduces the need for long-distance energy transportation through the electricity grid. To reduce congestion and supply electricity to a local demand of households, hydrogen storage can be an important flexibility option to bridge the seasonality gap associated with supply and a local electricity demand when external distribution capacity is limited. 

In this paper, we examine the problem of of the owner of a solar park with local hydrogen storage who needs to decide how much to store, sell to or buy from an external electricity grid throughout the year and can supply energy to local electricity demand by households. Furthermore, the solar energy production and local electricity demand are seasonal and there is uncertainty associated with solar electricity supply, electricity demand, and variable electricity prices in the external electricity market. We propose a Markov decision process formulation to the above problem to optimize the expected profits per year. We detail the optimal policies with regard to the period in which the actions take place (e.g., summer or winter). Moreover, we illustrate which actions are taken during overages and shortages throughout the year. We show how congestion levels and electrolyzer utilization are affected by conversion efficiency and strategic decisions such as the distribution capacity, storage capacity, and production capacity.

It is found that optimal policies are characterized by price thresholds which separate different types of actions. These include buying the maximum possible quantity, selling exactly overages or buying exact shortages, storing overages or obtaining shortages from storage, or selling the maximum amount possible. When distribution capacity is unconstrained, storage is not used for large periods of time. When distribution capacity is constrained, local congestion at the cable to which the solar park is connected is mostly caused by buying-related actions in winter, which are needed to cover potential future shortages. Under these conditions, increasing the level of storage capacity does not reduce congestion levels, because buying actions in winter remain necessary to cover shortages. For higher levels of distribution capacity, local congestion is mostly caused by selling-related actions of the overages in the summer. Counter-intuitively, local congestion increases for increased levels of storage capacity, because this enables increasing buying-related actions to prevent future shortages and exploiting price differences.

Mean profits are highly sensitive to the level of electrolyzer capacity and appear to increase linearly with capacity. Moreover, a lower electrolyzer utilization as a result of a large capacity is associated with higher profits than a low electrolyzer capacity with a higher utilization rate as a result of interacting with the electricity grid. This indicates that a high utilization rate of the electrolyzer is not necessarily an indication of increased profits. Hydrogen storage used to supply electricity does not lead to profits when distribution capacity is too small. Moreover, storage also does not aid in reducing local congestion at the connected cable when the associated distribution capacity is too small. This is because buying actions to prevent future shortages and benefit from price differences cause buying-related congestion at the cable connection. These actions are not aimed at reducing congestion, but at maximizing profits. Higher production capacities are associated with higher profits, even though this also causes higher congestion levels due to increased selling to the grid at times of abundant supply or high prices. 

The opportunities for future research are numerous and can be divided into two areas.
The first area is centered around extending the model that we present in this work. For instance, new concepts arise in which local demand not only consists of electricity but also of hydrogen and, for example, demand for heat. Additionally, one could investigate the potential correlation between production and demand, or physical properties of using hydrogen as an energy carrier. Other future approaches may be focused towards minimizing congestion instead of maximizing the storage owner's profits.

The other avenue for further research is the transition towards more strategic models. For instance, one may investigate the impact of multiple renewable energy systems with co-located storage facilities in grid-congestion issues. It would be interesting to research how the location of renewable energy systems in a grid can be optimized, with the aim of minimizing grid congestion. 

\subsection*{Acknowledgements}
This study was supported by The Netherlands Organisation for Scientific Research (NWO) with grant number 438-15-519. Declarations of interest: None.

\bibliographystyle{elsarticle-harv}
\bibliography{refs}






\end{document}